\documentclass[twocolumn,superscriptaddress,showpacs,preprintnumbers,amsmath,amssymb,prl]{revtex4}

\usepackage{graphicx}
\usepackage{dcolumn}
\usepackage{bm}

\newcommand{\sto}{SrTiO$_3$}
\newcommand{\qo}{$q_{0}$}

\newcommand{\Ispec}{$I_{spec}$}
\newcommand{\Idiff}{$I_{diff}$}
\newcommand{\Iisl}{$I_{isl}$}
\newcommand{\tspec}{$\tau_{spec}$}
\newcommand{\tisl}{$\tau_{isl}$}

\begin{document}

\preprint{Physical Review Letters}

\title{Measurements of Surface Diffusivity and Coarsening During Pulsed Laser Deposition}

\author{J. D. Ferguson}
 \affiliation{Department of Materials Science and Engineering, Cornell University, Ithaca, NY 14853, USA}
 \affiliation{Cornell Center for Materials Research, Cornell University, Ithaca, NY 14853, USA}
\author{G. Arikan}
 \affiliation{School of Applied and Engineering Physics, Cornell University, Ithaca, NY 14853, USA}
 \affiliation{Cornell Center for Materials Research, Cornell University, Ithaca, NY 14853, USA}
\author{D. S. Dale}
\affiliation{Cornell High Energy Synchrotron Source, Cornell University, Ithaca, NY 14853, USA}
\author{A. R. Woll}
\affiliation{Cornell High Energy Synchrotron Source, Cornell University, Ithaca, NY 14853, USA}
\author{J. D. Brock}
 \affiliation{School of Applied and Engineering Physics, Cornell University, Ithaca, NY 14853, USA}
 \affiliation{Cornell Center for Materials Research, Cornell University, Ithaca, NY 14853, USA}

\date{\today}

\begin{abstract}
Pulsed Laser Deposition (PLD) of homoepitaxial \sto$\langle001\rangle$ was studied with \textit{in-situ} x-ray specular reflectivity and surface diffuse x-ray scattering. Unlike prior reflecivity-based studies, these measurements access both the time- and the length-scales of the evolution of the surface morphogy during growth. In particular, we show that this technique allows direct measurements of the diffusivity for both inter- and intra-layer transport. Our results explicitly limit the possible role of island break-up, demonstrate the key roles played by nucleation and coarsening in PLD, and place an upper bound on the Ehrlich Schwoebel (ES) barrier for downhill diffusion.

\end{abstract}

\pacs{68.55.-a, 68.55.A-, 68.47.Gh, 81.15.Fg, 61.05.cf}

\maketitle

PLD presents an exceptional challenge for experimental and theoretical study due to its highly non-equilibrium nature, the vast range of time- and length-scales involved, and the complex stoichiometry of the materials system studied. Consequently, fundamental issues, such as the roles played by the pulsed nature and the kinetic energy of the deposit, remain unresolved\cite{Aziz:2008,Blank:1999,Fleet:2005,Tischler:2006,Willmott:2006}. System-specific kinetic properties are also difficult to obtain. For example, STM has revealed a rich variety of phenomena on \sto\ surfaces\cite{Lippmaa:1998,Marsh:2006}, but at time scales longer than those relevant to growth. In contrast, fast studies of PLD have typically employed electron\cite{Blank:1999,Lippmaa:2000} or x-ray\cite{Fleet:2005,Tischler:2006,Willmott:2006,Dale:2006,Fleet:2006} specular reflectivity. These studies have excellent time resolution, but are sensitive only to the average atomic-scale surface roughness\cite{Sinha:1988,Dale:2008}, and therefore provide an incomplete description of surface kinetics.

\begin{figure*}
\includegraphics{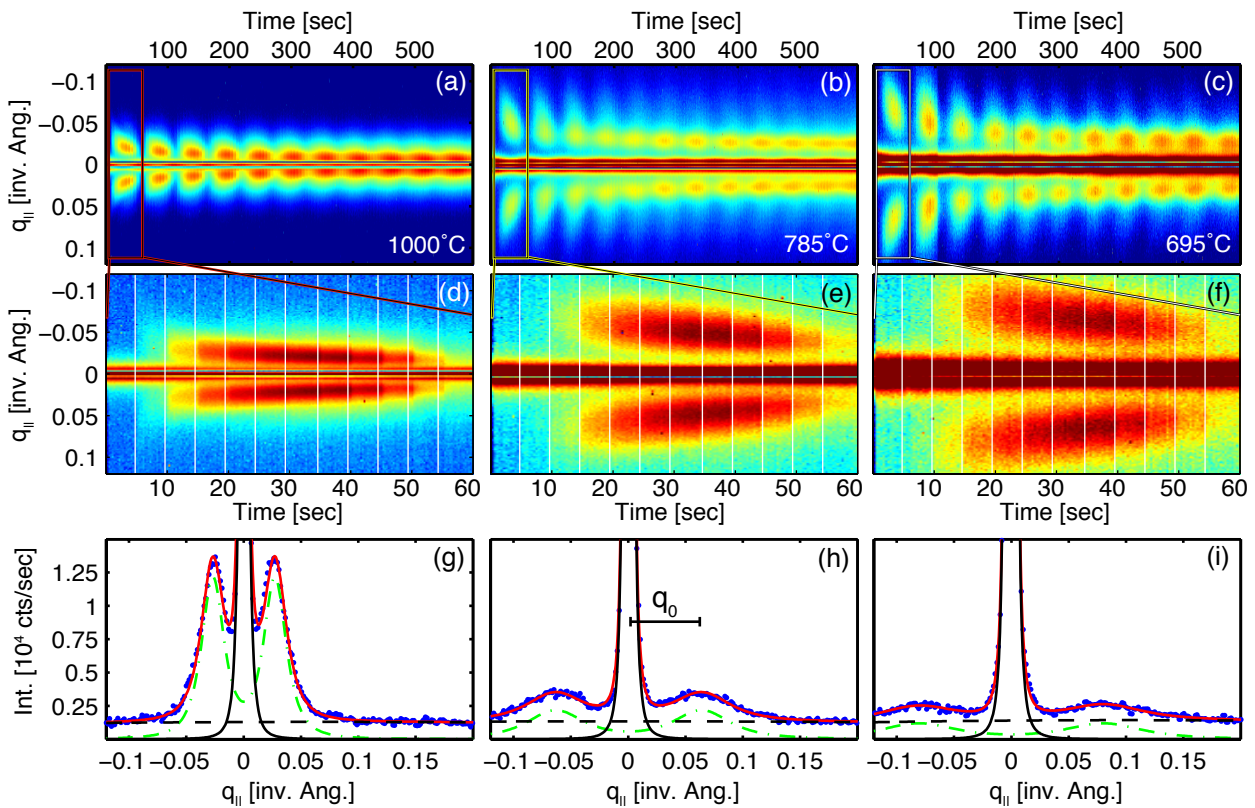}
\caption{\label{fig1}Diffuse x-ray scattering for the PLD of \sto$\langle001\rangle$. (a-c) Depositions of $\sim$11 ML at $1000^{\circ}$C, $785^{\circ}$C, $695^{\circ}$C, respectively. (d-f) The corresponding first ML. Vertical lines represent the laser pulses (first pulse at 5 s). (g-i) Scattering line shape at $t=16.5$ s for each temperature. $I_{fit}$ (red, solid) consists of \Idiff\ (green, dash-dot), \Ispec\ (black, solid) and $I_{bg}$ (black, dashed).}
\end{figure*}

In this Letter, we show that \textit{in-situ} x-ray diffuse scattering provides critical length scale information absent from x-ray reflectivity alone, at time scales sufficient to study PLD. The experimental details are given in Ref. \cite{Auxmat}. Figs.~\ref{fig1}(a-c) show false color images of the intensity of both the specular ($q_{||}=0$) and the surface diffuse scattering as a function of time and $q_{||}$, during the deposition of approximately 11 monolayers (ML) of unit cell step height at 3 temperatures.  As material is deposited on the surface, the specular intensity drops while diffuse lobes of scattering appear on both sides of the specular rod. These lobes are cuts through  ``Henzler rings'' arising from 2D islands on the surface\cite{Hahn:1980,Bartelt:1993},  as verified by \textit{ex-situ} atomic force microscopy (See Ref. \cite{Auxmat} Fig. 2).

At low layer coverage, $\theta$ ($0<\theta<0.4$ ML), the radius of these rings, \qo, is inversely proportional to the average island separation, $L_{isl}\approx2\pi/q_{0}$ \cite{Dulot:2003}. As more material is deposited, the intensity of the specular rod and the diffuse lobes oscillate out of phase with a period of 1 ML. Near layer completion ($0.7<\theta<1$ ML), \qo\ is a measure of the separation between holes rather than islands.

A conspicuous feature of Figs.~\ref{fig1}(a-c) is that increasing the substrate temperature results in a decrease in $q_{0}$, corresponding to a decrease in island density, as expected from classical nucleation theory\cite{Brune:1999}. A second feature of the data is that $q_{0}$ decreases with increasing layer number. This is a general feature of every data set we obtained, and reflects the growth surface's ``memory'' of underlying layers. If a new layer nucleates before layer completion, the remaining holes function as adatom sinks, suppressing the adatom density, thereby producing a smaller nucleation density. 

Figs.~\ref{fig1}(d-f) show an enlarged view of the 1st ML of growth from Figs.~\ref{fig1}(a-c). At $1000^{\circ}$C, diffuse scattering appears between the first and second pulses. At sufficiently lower temperatures ($\leq785^{\circ}$C), diffuse scattering is not visible until after the second pulse, indicating either delayed nucleation or intensity below our detection limit. 

To extract quantitative information, the x-ray data were fit to the sum of three independent components,
\begin{equation}
I_{fit}(q_{||})=I_{bg}+I_{spec}(q_{||})+I_{diff}(q_{||}\pm q_{0}).
\label{eq1}
\end{equation}
In this equation, $I_{bg}$ is a constant background, and \Ispec\ and \Idiff\ take the form:
\begin{equation}
f(x)=I_{0}/[1+\xi^{2}x^{2}]^{3/2},
\label{eq2}
\end{equation}
where $\xi$ is the correlation length. The parameters $I_{0}$ and $\xi$ each take on two values, associated with \Ispec\ and \Idiff. Eq. (\ref{eq2}) with $q_{0}=0$ corresponds to the scattering profile of a random distribution of islands\cite{Debye:1957,Bracewell:2000}. Figs.~\ref{fig1}(g-i) show the single frames from Figs.~\ref{fig1}(d-f) corresponding to $t=16.5$ s: the frame immediately following the third laser pulse. Also shown are the best fit to Eq. (\ref{eq1}) and its components. The agreement between the fitting function and the data is excellent, with a typical $\chi^{2}\approx1.3$.

\begin{figure}
\includegraphics{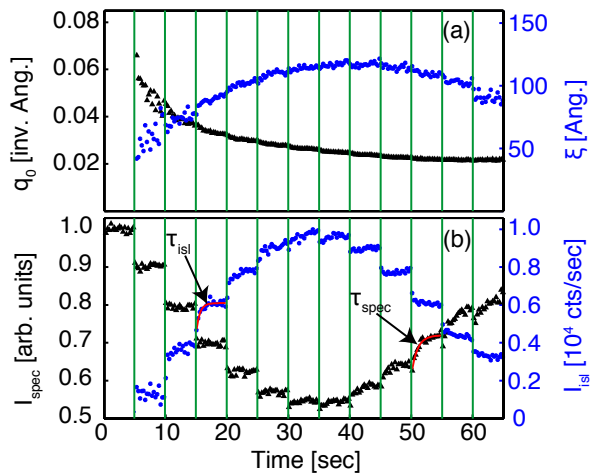}
\caption{\label{fig2} (a) The peak position of the diffuse lobes, $q_{0} (\triangle)$ and the correlation length, $\xi (\bullet)$ at $850^{\circ}$C are shown for the $1^{st}$ ML. Vertical lines represent laser pulses($1^{st}$ pulse at 5 s). (b) \Ispec\ $(\triangle)$ and \Idiff\ $(\bullet)$ are shown. The characteristic diffusion times, \tisl\ and \tspec\, are determined by fitting \Ispec\ and \Iisl.}
\end{figure}

Fig.~\ref{fig2}a shows the evolution of \qo\ and $\xi$ for the first monolayer at $850^{\circ}$C. Immediately following the first pulse, a diffuse peak is observed at $q_{0}=0.066\pm0.005$ \AA, indicating that some islands have nucleated. This value of \qo\ corresponds to an island density of $n_{x}=(1.0\pm0.1)\times10^{12}$ cm$^{-2}$ if a triangular lattice is assumed. A rising \qo\ immediately following the first pulse would signify nucleation of new islands from a supersaturation of adatoms. Instead, \qo\ decreases monotonically and continuously, indicating a steadily decreasing island density. This shows that some of the newly formed islands are disappearing, and thus that island coarsening\cite{Bartelt:1996,Wen:1996}, rather than nucleation, drives the evolution of \qo\ during this time. We observe similar coarsening for substrate temperatures as low as $695^{\circ}$C.

A key parameter in PLD growth is the decay time of the adatom supersaturation resulting from the pulse\cite{Jubert:2003}. Our diffuse scattering measurements are not directly sensitive to adatom supersaturation. Specifically, since they only extend only to $q_{max}=0.2$ \AA $^{-1}$ (see Figs.~\ref{fig1}(g-i)), they are insensitive to lateral correlations smaller than $\approx2\pi/q_{max}\approx31$ \AA, such as adatoms or very small islands. However, it is easily shown that, if the coverage and specular intensity are both constant, the total diffuse scattering intensity is also constant\footnote{By integrating Eq. (6) in Ref. \cite{Sinha:1996} over $q_{||}$, it is seen that both \Ispec\ and \Idiff\ depend only on $h(q_{z})$, the Fourier transform of the vertical height distribution.}. We therefore write the total in-plane surface scattering as $I_{tot}=I_{spec}+I_{isl}+I_{sm}$, where $I_{sm}$ accounts for features not captured by our detector. \Iisl\ is equal to \Idiff\, from Eq. (\ref{eq1}), integrated over the $q_z$ plane:
\begin{equation}
I_{isl}=2\pi I_{0}(q_{0}/\xi)[1+\sqrt{1+(\xi q_{0})^{-2}}],
\label{eq3}
\end{equation}
and is associated with the total diffuse scattering due to large islands, i.e. the islands separated by $>2\pi/q_{max}$. When the specular intensity between pulses is constant, such as at low coverage in Fig. 2b, a time-dependent \Iisl\ corresponds to mass transfer between small features and the characteristic large islands that give rise to \Iisl.

The specular intensity, \Ispec, and total diffuse intensity \Iisl\ for an $850^{\circ}$C deposition are shown in Fig.~\ref{fig2}b. Apart from the jumps in \Ispec\ associated with each deposition pulse, we observe two, distinct slower changes occurring between pulses. The first is a change in \Ispec\ that occurs near monolayer completion and has been studied previously\cite{Fleet:2005,Blank:1999,Tischler:2006,Willmott:2006,Lippmaa:2000}. The second slow change, which manifests in \Iisl\ and has not previously been reported, occurs at low coverage. After the third laser pulse, the rise in \Iisl\ lags behind the fast drop in \Ispec. As discussed above, this delay indicates an increase in the amount of material in large islands. Moreover, since \Ispec\ is constant during this time, this mass transfer corresponds solely to intralayer transport. 

\begin{figure}
\includegraphics{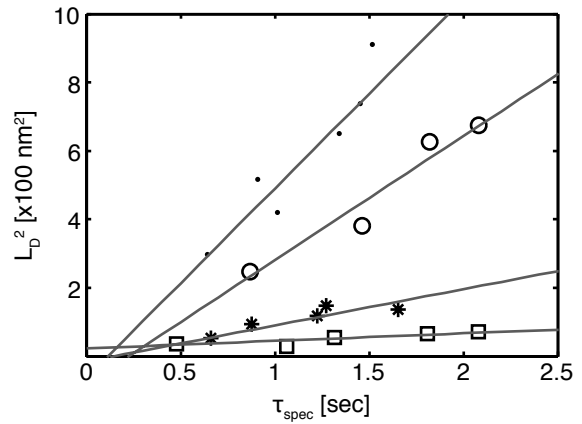}
\caption{\label{fig3}Length scale for diffusion, $L_{D}^{2}$ vs. \tspec\ for $1000^{\circ}$C $(\bullet)$, $850^{\circ}$C $(\circ)$, $785^{\circ}$, $(\ast)$, and $695^{\circ}$C $(\Box)$. The linear relationship shows that diffusion is the rate limiting process.}
\end{figure}

The relaxation kinetics described above can be quantified by fitting \Ispec\ at high $\theta$, and \Iisl\ at low $\theta$ to a simple exponential with characteristic relaxation times \tspec\ and \tisl. However, the physical process or processes giving rise to these time constants cannot be determined from Fig.~\ref{fig2} alone. For example, the diffusing species may come from pre-existing islands; therefore, both \tspec\ and \tisl\ may be determined by either the rate of adatom detachment or the rate of surface diffusion. If present, an ES barrier for downhill diffusion would also contribute to \tspec. We are able to resolve this ambiguity by examining the relationship between \tspec\ and \qo\ obtained for different layers in a single growth, exploiting the fact that \qo\ decreases with increasing layer number. If diffusion is indeed the rate-limiting process determining \tspec\, and if the average diffusion length, $L_{D}$, is determined by \qo\, then the Einstein relation, $L_{D}^{2}=4D\tau$ applies\cite{Gomer:1990}. We associate each \qo\ with an approximate diffusion length $L_{D}=L_{isl}/2=\pi/q_{0}$, (approximately half the distance between hole centers), and plot $L_{D}^{2}$ vs. \tspec\ obtained from approximately the same coverage, e.g. $\theta\approx0.8,1.8,2.8$, for several different layers in Fig~\ref{fig3}. A clear linear relationship is observed, so that we may associate the slope in Fig.~\ref{fig3} with the diffusivity, $D$. We also assign \tisl\ to diffusion-limited transport, since only a subset of the processes responsible for \tspec\ are involved.

\begin{figure}
\includegraphics{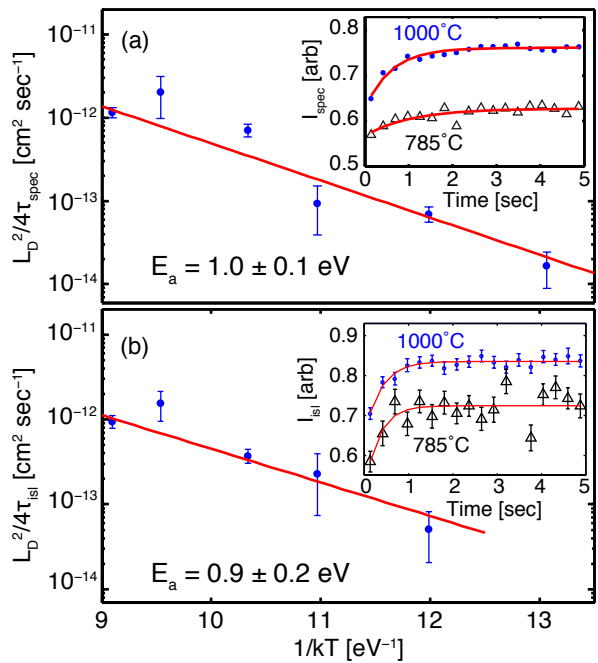}
\caption{\label{fig4} (a) Arrhenius behavior of the diffusivity at $\theta\approx0.8$ML. (inset) \tspec\ is obtained from the specular relaxation at high coverage, during inter-layer transport. (b). Diffusivity at at $\theta\approx0.25$ ML. (inset) \tisl\ is determined by fitting the time evolution of \Iisl.}
\end{figure}

Figs.~\ref{fig4}(a,b) show Arrhenius plots of $D$ obtained from the analysis of \tspec\ and \tisl\ for the first ML. The best-fit lines are shown, corresponding to activation energies of $E_{a}=1.0\pm0.1$ eV and $E_{a}=0.9\pm0.2$ eV for inter- and intra-layer transport, respectively. The difference in these energies, $0.1\pm0.22$ eV, is a direct measure of the ES barrier. Remarkably, these data sets yield not only the same slopes (within experimental error) but also the same values of diffusivity throughout the temperature range studied, suggesting that the ES barrier is negligible. We thus combine the data in Figs.~\ref{fig4}(a-b) to give the single result $D=D_{0}\exp(-E_{a} / k_{B}T)$, with $D_{0}=10^{-8\pm1}$cm$^{2}$ s$^{-1}$ and $E_{a}=0.97\pm0.07$ eV. The determination of both $D_{0}$ and $E_{a}$ through diffraction-based measurements alone represents a principle result of this work. 

The value of $E_{a}$ reported here is larger than two values, $0.48\pm0.05$ eV and $0.6\pm0.2$ eV, previously reported in the literature\cite{Blank:1999,Fleet:2006}. In these reports, $E_{a}$ was obtained from the temperature dependence of \tspec\, implicitly assuming a constant length scale. The effect of this assumption on the determination of $E_{a}$ is made explicit by writing the temperature dependence of the length scale in Arrhenius form, $L_{D}=L_{0}\exp(-E_{L}/k_{B}T)$, and rewriting the Einstein relation:
\begin{equation}
\tau_{spec}=(L_{0}^{2}/4D_{0})\exp[(E_{a}-2E_{L})/k_{B}T]
\label{eq5}
\end{equation}

Eq. (\ref{eq5}) shows that the activation energy measured from \tspec\ alone underestimates the activation barrier for diffusion, $E_{a}$, by $2E_{L}$. We note that our value of $E_{a}=0.97\pm0.07$ eV is very close to that of $1.2\pm0.1$ eV measured for diffusion of TiO$_{x}$ ``diline'' units on a reconstructed \sto\ surface\cite{Marsh:2006}.

Our results provide new insight into the possibility of energetic mechanisms promoting smooth growth in complex oxide PLD. One such proposed mechanism is island breakup, in which energetic impinging material breaks up existing islands, delaying second-layer nucleation. Island-breakup has previously been observed in simulations of metal/metal epitaxy\cite{Jacobsen:1998,Pomeroy:2002} and was recently invoked\cite{Willmott:2006} to explain experimental results of PLD of La$_{1-x}$Sr$_x$MnO$_3$ on \sto. Specifically, Ref.\cite{Willmott:2006} suggests that island break-up produces an increasing island density when $\theta<0.5$ ML. Although the system studied here is not precisely the same as in Ref.\cite{Willmott:2006}, Fig.~\ref{fig2}a demonstrates that the island density monotonically decreases with $\theta$ from the earliest moments after nucleation. Island breakup could also manifest in our measurement as a decrease in \Iisl\ as mass is transferred from large islands to smaller species without changing $q_{0}$. However, we do not observe such a decrease. Thus, island coarsening overwhelms any possible effect of island break-up.

A second proposed non-thermal smoothing mechanism suggested by prior experimental work on complex-oxide PLD, is enhanced downhill transport\cite{Blank:1999,Fleet:2005,Tischler:2006,Willmott:2006}. The experimental basis for this suggestion is the observation, based on specular reflectivity, that downhill transport occurs on two widely separated time scales\cite{Fleet:2005,Tischler:2006}. Our observation, that island nucleation occurs quickly, followed by coarsening, suggests an alternate origin of these two time scales. Specifically, it is possible that the mobile species responsible for slow downhill transport consists of material that detaches from islands. This material need not be chemically identical with the species arriving from the plume. Interestingly, we note that the prefactor reported here, $D_{0}=10^{-8\pm1}$cm$^{2}$ s$^{-1}$, is five orders of magnitude lower than typical experimental and theoretical value for metal and semiconductor systems\cite{Kaxiras:1994}. Similar diminished prefactors have previously been associated with correlated motion involving multiple atoms\cite{Kaxiras:1994}. Here, it might be associated with stoichiometric mass transfer of Sr-containing and Ti-containing species.

In summary, we have presented time-resolved x-ray reflectivity and diffuse scattering measurements obtained during PLD. Our results constitute direct observations of island nucleation as little as $200$ ms after the pulse, and direct evidence of island coarsening occurring between laser pulses for temperatures as low as $695^{\circ}$C. Quantitative analysis of our results allow us to independently estimate the inter- and intra-layer diffusivity (prefactor and activation barrier) of mobile species between pulses and to place an upper bound on the ES barrier. Our measurements significantly impact prior estimates of the thermal diffusivity involved in \sto\ growth, and place specific constraints on energetic smoothing mechanisms that have been proposed to occur during complex oxide PLD.

\begin{acknowledgments}
We thank J. Blakely, Y. Kim, H. Wang, D. Muller and M. Tate. This research was conducted in part at the Cornell High Energy Synchrotron Source and was supported by NSF (DMR-0317729 and DMR-0225180).
\end{acknowledgments}

\section{\label{sec:level1}Experimental Details.} 

\begin{figure}
\includegraphics{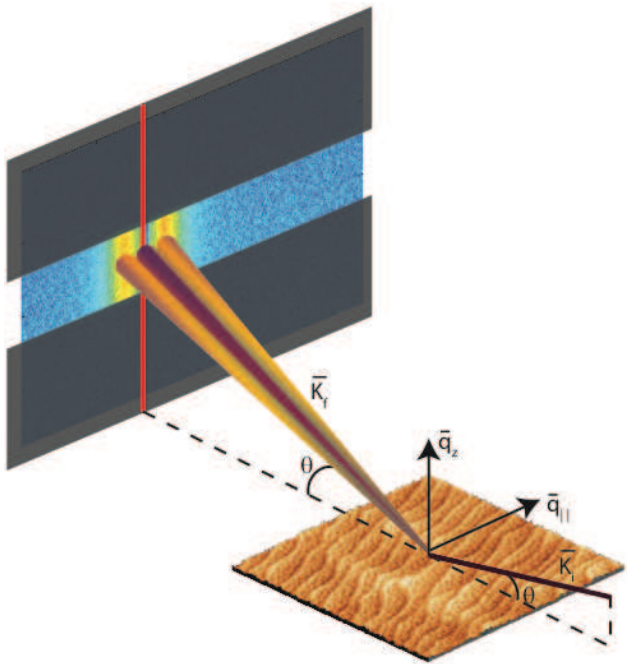}
\caption{\label{setup}Diffuse Scattering Diagram.  A 10 keV x-ray beam reflects off the surface with a specular (purple) and diffuse scattering (gold) which is detected with a CCD detector. The value of $q_{||}$ measured is directly related to the island correlations on the surface.}
\end{figure}

The x-ray measurements were performed using a custom PLD/x-ray diffraction system installed in the G3 hutch at the Cornell High Energy Synchrotron Source. Auxiliary Fig.\ref{setup} illustrates the experimental geometry. The surface shown is an atomic force microscopy (AFM) image taken after the deposition performed in Fig. 1(a) of the manuscript where $\theta=11.3$ ML. As suggested by the figure, steps on the sample resulting from the surface miscut are always aligned perpendicular to the incident beam. The x-ray beam produces scattering in the $q_{||}$ direction that is directly related to the island correlations on the surface. For these experiments, a monochromatic ($\Delta$E/E=1\%) 10.0 keV x-ray beam with $8\times10^{13}$  photons/sec/mm$^{2}$ was slit down to produce a 1.0 mm $\times$ 0.5 mm beam at the sample. The specular component of the reflected signal was attenuated using a tin absorber to prevent detector saturation. To optimize the signal-to-background ratio of the diffuse scattering, the experiment was performed near the $L=0.275$ r.l.u. position on the crystal truncation rod. The in-plane diffuse surface scattering was monitored using a CCD area detector operating as a linear detector in Òstreak mode.Ó The time resolution of the experiment is limited by both the readout time of the detector ($\approx78$ ms) and by the incident x-ray flux ($100-200$ ms). For these growth conditions, we are able to collect 18 images between laser pulses. 

The depositions were performed by laser ablating a single crystal \sto\ target using a 100 MW/cm$^{2}$ KrF excimer laser (248 nm). The target is located 6 cm from the substrate. The area of the laser spot on the target was approximately 3.7 mm$^{2}$ with a fluence of 1.9 J/cm$^{2}$. This configuration deposited $\approx$0.09 ML/pulse at a laser repetition rate of 0.2 Hz, with a $2\times10^{-4}$ Torr partial pressure of O$_{2}$. The substrate temperature was measured using an optical pyrometer ($\lambda=4.8-5.3 \mu$m, emissivity=0.8). The substrate preparation procedure employed\cite{Koster:1998,Kawasaki:1994} produced a TiO$_{2}$ terminated surface, and AFM confirmed the presence of single unit cell high steps separating large atomically flat terraces.

\begin{figure}
\includegraphics{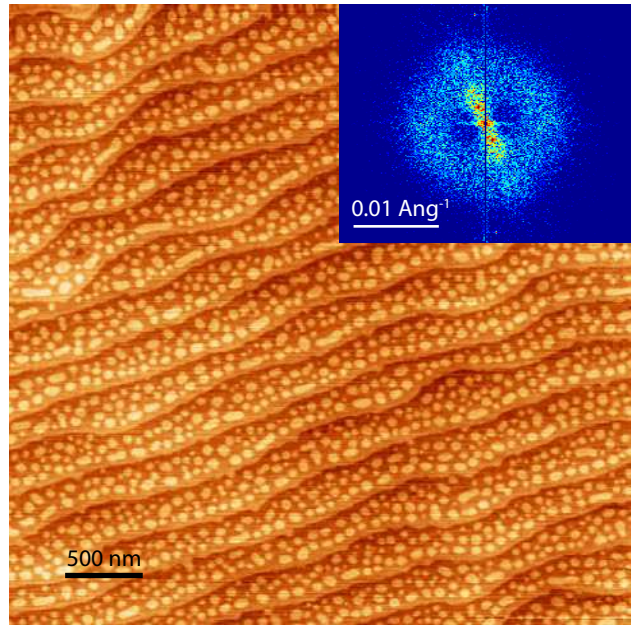}
\caption{\label{afm}AFM after 11.3 ML Deposition: Post deposition AFM of the 1000ûC growth reveals unit cell high islands on the surface. The diffuse ring present in the fast fourier transform (FFT) shows that the islands are correlated (inset).}
\end{figure}

The diffuse scattering peaks are the direct result of single unit cell high islands, which was confirmed by ex-situ AFM. An AFM image after the deposition at 1000ûC, shown in Fig. 1(a) of the manuscript, is given in Auxiliary Fig.\ref{afm} confirming the presence of islands on the surface. The fast Fourier transform (FFT) of the image is shown in the inset. The presence of the diffuse Henzler ring is a direct result of the correlated islands on the surface. Our experimental detected x-ray intensity is a cut through these Henzler rings. Coarsening, during the post deposition cool-down ($\approx3$ hours), has caused the increase of the measured length scale of the AFM relative to the x-ray data.

\end{document}